\documentclass[aps,nofootinbib,preprintnumbers,reprint,onecolumn,floatfix]{revtex4-1}
\usepackage{graphicx}
\usepackage{amssymb,amsmath,mathrsfs}
\usepackage{cancel}
\usepackage{subfig}
\usepackage{hyperref}
\usepackage{verbatim}
\usepackage{xspace}

\def\bea{\begin{eqnarray}}
\def\eea{\end{eqnarray}}
\newcommand{\Mpl}{M_\mathrm{Pl}}
\newcommand{\TeV}{\mathrm{TeV}}
\newcommand{\GeV}{\mathrm{GeV}}

\newcommand{\TRH}{T_\mathrm{RH}}

\newcommand{\beq}{\begin{equation}}
\newcommand{\eeq}{\end{equation}}

\newcommand{\hc}{\mathrm{h.c.}}
\newcommand{\neqd}{n^{\mathrm{eq}}}

\newcommand{\Gamtot}{\Gamma_{\varphi}}
\newcommand{\B}{$B$}
\newcommand{\C}{$C$}
\newcommand{\CP}{$CP$\xspace}
\newcommand{\CPT}{$CPT$\xspace}
\newcommand{\mB}{m_{\psi_B}}
\newcommand{\mD}{m_{\psi_D}}
\newcommand{\mDB}{m_{\Phi_{DB}}}

\DeclareMathOperator\Ei{Ei}
\begin{document}
\title{Particle Asymmetries from Quantum Statistics}
\author{Nikita Blinov}
\affiliation{SLAC National Accelerator Laboratory, 2575 Sand Hill Road, Menlo Park, CA, 94025, USA}
\author{Anson Hook}
\affiliation{Stanford Institute for Theoretical Physics, Stanford University, Stanford, CA 94305, USA}
\date{\today}
\preprint{SLAC-PUB-16936}
\begin{abstract}
We consider a class of baryogenesis models where the Lagrangian in the visible
sector is Charge-Parity ($CP$) invariant and a baryon asymmetry is produced only
when quantum statistics is taken into account. The $CP$ symmetry is broken by
matter effects, namely the assumption that the primordial plasma contains
another asymmetric species, such as dark matter.  Out-of-equilibrium baryon
number violating decays can then generate an asymmetry through Bose enhancement
and/or Pauli blocking of certain decay channels.
\end{abstract}
\maketitle

\setlength{\parindent}{4ex}

\section{Introduction}

The origin of the observed excess of matter over anti-matter in the universe
is one of the fundamental questions in particle physics~\cite{Cohen:1997ac}. A dynamical 
explanation must satisfy Sakharov's three criteria: 
violation of baryon number (\B), violation of charge (\C) and charge-parity (\CP) invariance, and a departure from
thermal equilibrium~\cite{Sakharov:1967dj}. In many proposed scenarios \CP violation is intrinsically tied to the departure from 
thermal equilibrium. For example, in electroweak baryogenesis \CP-violating 
interactions with the advancing bubble wall (which drives 
plasma just outside out of equilibrium) are responsible for the generation 
of a chiral asymmetry that is then reprocessed into baryons~\cite{Kuzmin:1985mm,Morrissey:2012db}. In the standard 
out-of-equilibrium decay scenarios like GUT baryogenesis and leptogenesis, 
the couplings of the decaying particle violate \CP, allowing for an asymmetry 
to be created (see Refs.~\cite{Nanopoulos:1979gx,Fukugita:1986hr} and the reviews~\cite{Riotto:1998bt,Buchmuller:2005eh,Davidson:2008bu}). 
In this paper, we explore a class of models 
where \CP violation and the departure from thermal equilibrium are 
disentangled. We consider scenarios where an existing asymmetric particle 
density biases an otherwise \CP-conserving process through the 
effects of quantum statistics, i.e. Pauli blocking and Bose 
enhancement, resulting in baryon number production.
 
Charge-Parity violation can occur in ways that are not seen in the visible
sector Lagrangian.  After all, the prevalence of baryons over anti-baryons is
itself a violation of \CP.  Thus, it is clear that aside from a fundamental
parameter in the Lagrangian, \CP can also be violated by matter effects, 
i.e. by any pre-existing charge densities. This matter-induced \CP
violation has the distinct advantage of typically being
testable if the charge corresponding to the asymmetry is conserved until late
times. For example, this is the case for asymmetric dark matter (DM). The
asymmetry and stability on cosmological time scales guarantee that there is a particle currently present in the universe
that has a \CP-violating number abundance.  Note that we make a distinction
between the \CP breaking number abundances and the \CP breaking Lagrangian
parameters that they often come from (for some exceptions see
Ref.~\cite{Linde:1985gh, Hertzberg:2013mba, Unwin:2015wya, Hook:2015foa}). We
will consider the case where physics in the visible sector is \CP-preserving (up
to the Standard Model (SM) Cabibbo-Kobayashi-Maskawa phase) and \CP violation in the early universe results from the
presence of an asymmetric particle population rather than \CP
violation that might have caused their production.

There are several observed particles that can break \CP with their number
densities and therefore can be used to implement baryogenesis. Photons and gravitons can 
be chiral and thus break \CP. For example, chiral magnetic fields in 
the early universe can generate a $B+L$ asymmetry via the weak anomaly~\cite{Giovannini:1997eg,
Anber:2015yca, Fujita:2016igl,Kamada:2016cnb}, while chiral gravitational waves can 
source a $B-L$ asymmetry through the gravitational anomaly~\cite{Alexander:2004us}.

A dark matter asymmetry can also source the creation 
of baryon number. There are many ways to achieve this. The asymmetric dark matter (ADM)
paradigm postulates that dark matter \emph{carries} baryon number.  Thus an
asymmetry in DM entails an asymmetry in baryons; it is communicated to the SM
via a transfer operator~\cite{Zurek:2013wia}.  However, in this
case, dark matter is not the source of \CP violation, but rather a 
hidden reservoir of baryons (or anti-baryons~\cite{Davoudiasl:2010am}). 
The alternative we consider in this work utilizes the
fact that the DM number density $J_D^0 \ne 0$ breaks \CP, which 
can be used to generate baryon number from an otherwise \CP preserving decay. 
The existence of a chemical potential splits the energy levels of
particles and anti-particles. As a result, the \CPT symmetry is broken in
this non-vacuum background. This can also be seen from the fact that $J^0_D$ is \CPT odd and non-zero in the presence of a dark asymmetry. 
Thus, \CPT breaking allows for baryon asymmetry production at tree level without any interference effects.
The use of a \CPT-violating background is similar in spirit to models of spontaneous 
baryogenesis~\cite{Cohen:1987vi,Cohen:1988kt}~\footnote{In the model given in Eq.~\ref{eq:bose_model}, one can explicitly see the difference between spontaneous baryogenesis and the models we consider in this paper.  If the interactions were in thermal equilibrium, then there would be no baryon number generated.}.
Another example of this effect is the use of $J_D^0$ to generate a \CP violating coupling
in the Lagrangian much in the same way that the Higgs vacuum expectation value
allows for one to write a $SU(2)_W$ breaking Lagrangian
coupling~\cite{D'Agnolo:2015pha}.

In this paper, we use quantum statistics to transmit the \CP violation
from the dark sector to the SM.  As a simple example of this mechanism in
action, consider the out-of-equilibrium decays of a real scalar $\varphi$ with the interaction
\bea
\mathcal{L} \supset \frac{1}{\Lambda} \varphi \psi_B \psi_D \phi^\dagger_D
+ \hc
\label{eq:bose_model}
\eea
where $\psi_B$ is a fermion that carries baryon number and $\psi_D$
($\phi_D$) is a fermion (scalar) that carries a $U(1)_D$ dark quantum
number.  This interaction gives rise to two decay channels for $\varphi$, 
\bea
\label{eq: baryon}
\varphi &\rightarrow& \psi_B \psi_D \phi^\dagger_D \\
\label{eq: anti-baryon}
\varphi &\rightarrow& \psi_B^\dagger \psi_D^\dagger \phi_D.
\eea
The decays of $\varphi$ violate baryon
number but preserve dark matter number. In the absence of any \CP violation,
Lagrangian or otherwise, these two decays have equal probabilities so that no
baryon number asymmetry is generated. Now suppose that the existing 
DM density is asymmetric: the plasma contains more $\psi_D^\dagger$ ($\phi_D^\dagger$) 
than $\psi_D$
($\phi_D$). Given the simple set-up above, this is 
the only source of \CP violation. At finite density and temperature,
the $\varphi$ decay rate includes the effects of Pauli blocking 
and Bose enhancement due to the existence the final state particles in the plasma. As a result 
the channel~\ref{eq: baryon} is preferred over~\ref{eq: anti-baryon}, so 
the decays produce a baryon number asymmetry. In the limit of a large dark
matter asymmetry, the anti-baryon channel~\ref{eq: anti-baryon} can be completely forbidden. We
see that both boson and fermion statistics generate an asymmetry with the same sign 
at tree level. As we will show in Sec.~\ref{Sec: bose}, the
effect of Bose enhancement is significantly larger than Pauli exclusion for
this model.

In the scenarios we consider the baryon asymmetry is roughly bounded from above by the dark sector asymmetry. 
If this asymmetry persists to late times, 
the dark sector particle must be lighter than baryons since $\Omega_{\mathrm{cdm}}/\Omega_{b}\sim 5$. 
However, if these states eventually decay, their mass is not constrained. In what follows, we refer 
to the asymmetric dark sector states as DM, even if they are unstable on cosmological timescales and 
do not comprise the entire DM density of the universe today.

Standard baryogenesis via out-of-equilibrium decay is an ``infra-red dominated'' process, in which
the decays of the particle and the desired asymmetry are generated at the same time 
$t\sim H^{-1} \sim \Gamma_\varphi^{-1}$, where $\Gamma_\varphi$ is the $\varphi$ decay rate.  
The small fraction of decays that occur when $\Gamma_\varphi t \ll 1$ is 
irrelevant for the production of the asymmetry. This intuition rests on the assumption that the 
decay asymmetry does not depend on temperature. In contrast, in the models where quantum statistics is responsible for the generation of the asymmetry, we find important temperature 
dependence. As we describe in
the following sections, this causes the majority of the asymmetry to be produced at early times, 
well before $t \sim \Gamma_\varphi^{-1}$.

This paper is organized as follows. In Section~\ref{Sec: bose}, we consider the
model of Eq.~\ref{eq:bose_model} in detail. We show  
numerically and analytically that Bose enhancement of 
individual decay channels can result in a baryon asymmetry parametrically of the same size as the dark matter asymmetry. 
Surprisingly, we find that for certain parameters the baryon asymmetry can be larger by $\mathcal{O}(1)$ factors.  In
Section~\ref{Sec: pauli}, we present a second model where Pauli exclusion rather
than Bose enhancement is the dominant effect responsible for generating a large
asymmetry in the visible sector.  
We discuss our results and conclude in Section~\ref{Sec: conclusion}.

\section{Asymmetries through Bose Enhancement} \label{Sec: bose}
In this section, we examine the model presented in the introduction.  
We consider the Lagrangian
\bea
\mathcal{L} \supset \left(\frac{1}{\Lambda}\varphi \psi_B \psi_D \phi^\dagger_D - m_{\psi_D} \psi_D \psi_D^c - m_{\psi_B} \psi_B \psi_B^c + \hc \right) -m_{\phi_D}^2 |\phi_D|^2 .
\label{eq:bose_model_full}
\eea
The fermion $\psi_B$ carries baryon number
$B$ and its interactions with $\varphi$ break $B$. We assume that 
an asymmetry in $\psi_B$ can be converted into a 
SM baryon asymmetry through, e.g., the neutron portal 
\beq
\mathcal{L} \supset \frac{1}{\Lambda^2}\psi_B u^c_R d^c_R d^c_R + \hc
\label{eq:neutron_portal}
\eeq
We 
neglected the allowed interaction $\varphi \psi_B \psi_D^c \phi_D/\Lambda'$. 
If included, it would not
qualitatively change the results as long as $\Lambda' \ne \Lambda$.

Chemical 
equilibrium among the dark sector states $\psi_D$ and $\phi_D$ can 
be maintained with the inclusion of additional states that can mediate the 
reaction $\psi_D \psi_D \leftrightarrow \phi_D\phi_D$. This process 
can occur through an $s$-channel exchange of a $U(1)_D$-charged scalar $\Phi$ with interactions
\beq
\Phi^\dagger \phi_D \phi_D  + \Phi^\dagger \psi_D \psi_D + \Phi \psi_D^c \psi_D^c +\hc,
\eeq
or through a $t$-channel fermion 
mediator $\chi$ coupling to DM via
\beq
\phi_D^\dagger \psi_D \chi + \phi_D \psi_D^c \chi + \hc 
\eeq
In what follows we remain agnostic to the origin of the chemical equilibriation of DM 
with itself.  The first term in Eq.~\ref{eq:bose_model_full} combined with one of the equilibriation 
mechanisms above generates a Majorana mass for
$\psi_B$, which is small for parameters of interest (where $\varphi$ decays 
out of equilibrium) and will be ignored.   

We imagine that the scalar $\varphi$ decays far out of equilibrium when
the universe is already populated by asymmetric dark matter. At finite temperature the
rate for a single decay channel is 
\beq
\Gamma(\varphi\rightarrow \psi_B \psi_D \varphi_D^\dagger) = 
\frac{1}{2M_\varphi} \int d\Phi_3 |\mathcal{M}|^2 
(1 + f_{\phi_D})
(1 - f_{\psi_D})
(1 - f_{\psi_B}),
\label{eq:basic_decay_rate}
\eeq
where $\mathcal{M}$ is the decay matrix element, $d\Phi_3$ is the three-body
phase space volume element.  The distribution functions $f_i$ have their
equilibrium Bose-Einstein or Fermi-Dirac forms 
\beq
f_{\mathrm{BE},\mathrm{FD}} = \left[\exp\left({\frac{E-\mu}{T}}\right)\mp1\right]^{-1},
\eeq
for bosons and fermions, respectively. The sign of the chemical potentials is
reversed for anti-particle distributions. Chemical potentials are related to
particle density asymmetries via
\beq
\Delta n_i = n_i - n_{\bar{i}} = g_i \int \frac{d^3 p}{(2 \pi)^3} \left[f(E,\mu) - f(E,-\mu)\right],
\label{eq:asym_from_chem_pot}
\eeq
where $g_i$ is the number of internal degrees of freedom of species $i$.
We require that $ |\mu_{\phi_D}| < m_{\phi_D}$ in order to avoid $\phi_D$
getting a vacuum expectation value.

The product of statistical factors in Eq.~\ref{eq:basic_decay_rate} 
encodes stimulated emission (Bose enhancement) and Pauli blocking 
by the particles already present in the bath.
Note that their effects are large in the region of phase space where the final
state particles are produced with energy less than $T$. As we will be taking
$M_\varphi \gg T$, when one particle has energy $\lesssim T$, the other two
will have large energies of order $M_\varphi$.

\begin{figure}
  {\centering
  \includegraphics[width=0.45\textwidth]{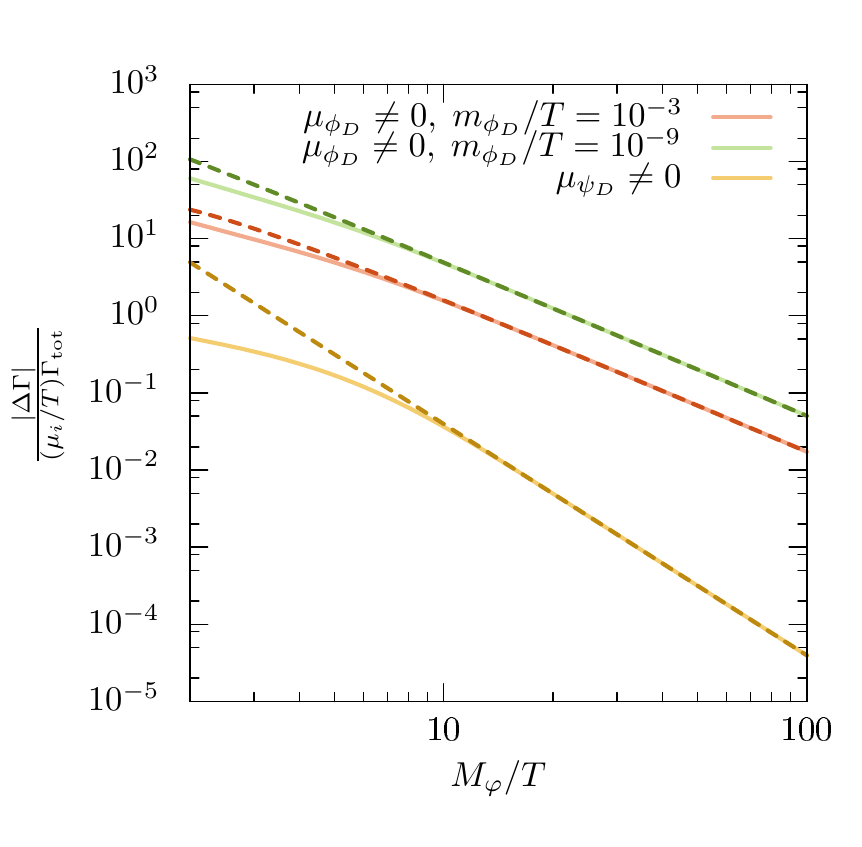}
  \includegraphics[width=0.45\textwidth]{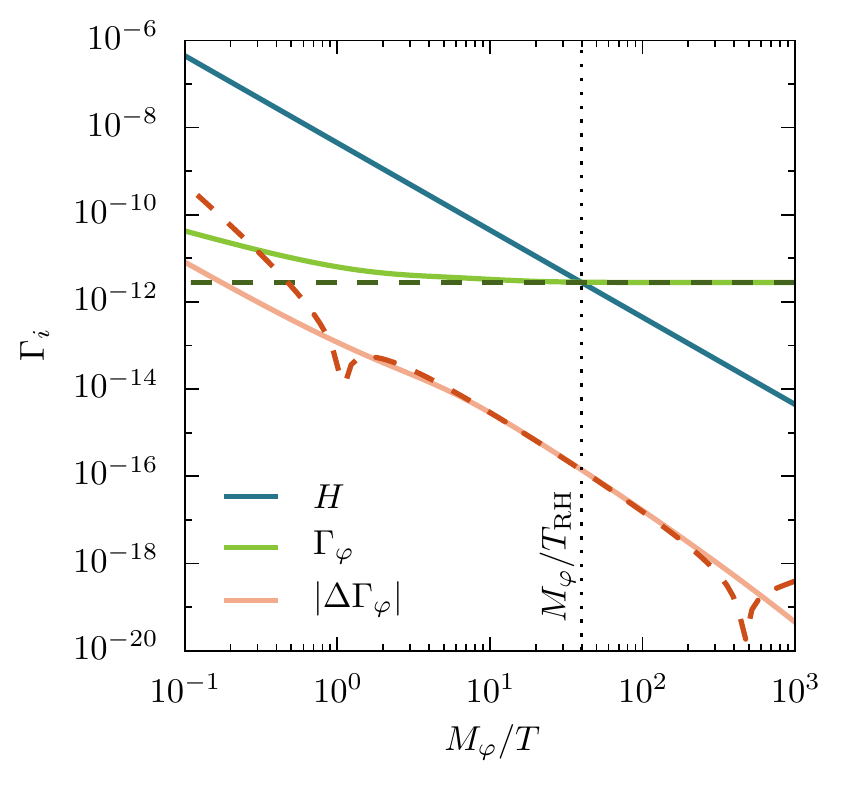}
}
  \caption{
    Decay widths as function of $M_\varphi/T$. The left plot compares 
    numerical computations (solid lines) to the analytic approximations (dashed lines) of the decay asymmetry discussed in the text 
  for fixed value of $m_{\phi_D}/T$. Contributions from Bose enhancement due to $\phi_D$
  and Pauli blocking due to $\psi_D$ are isolated by choosing $\mu_{\phi_D}\neq 0$ 
  and $\mu_{\psi_D} \neq 0$ one at a time. The chemical potentials are assumed to be small, such that 
  the resulting $\Delta \Gamma$ is linear in $\mu_i/T$. The right plot compares the various 
  physical rates to Hubble for fixed masses $m_{\phi_D} = m_{\psi_D} = m_{\psi_B} =300\;\GeV$, $M_\varphi = 100\;\TeV$ 
  and $\Delta n_{\phi_D}/s = 10^{-5}$. As before, the solid lines are numerical calculations 
  while the dashed lines correspond to analytic approximations. 
  The dips in $\Delta \Gamma$ show where these approximations break down, namely where 
  $M_\varphi/T\sim 1$ and $m_{\phi_D}/T \sim 1$.
    \label{fig:decay_asym_three_body}}
\end{figure}

The total width of $\varphi$ at leading order in $T/M_\varphi$ is given by
\beq
\Gamtot = \Gamma(\varphi\rightarrow \psi_B \psi_D \phi_D^\dagger) + \Gamma(\varphi\rightarrow \bar \psi_B \bar \psi_D \phi_D)
= \frac{1}{768\pi^3} \frac{M_\varphi^3}{\Lambda^2}. 
\eeq
The dependence on chemical potentials of final state particles and the effects
of the statistical factors enter at higher order in $T/M_\varphi$.  In
the right panel of Fig.~\ref{fig:decay_asym_three_body}, we show the numerical and analytic results for
the total width.
As long as $M_\varphi \gtrsim 3 T$, the analytic estimate provides a good
approximation for the total width.  
The decay width determines the number density of
$\varphi$ through the Boltzmann equation
\beq
\dot n_\varphi + 3H n_\varphi = - \Gamtot n_\varphi.
\label{eq:phi_boltzmann}
\eeq
The out-of-equilibrium assumption ensures that inverse decays are not important.

As $\varphi$ decays, an asymmetry $\Delta n_{\psi_B} \ne 0$ can be generated because 
the rates for the two decay channels of $\varphi$ are not equal when the DM 
is asymmetric, i.e. when there are non-zero chemical potentials for $\phi_D$ and $\psi_D$. 
The resulting production of baryon number is governed by 
\beq
\Delta \dot n_{\psi_B} + 3H \Delta n_{\psi_B} = \Delta \Gamma n_\varphi, 
\label{eq:baryon_boltzmann}
\eeq
where the decay asymmetry 
\beq
\Delta\Gamma =
\Gamma(\varphi\rightarrow \psi_B \psi_D \phi_D^\dagger) - \Gamma(\varphi\rightarrow \bar \psi_B \bar \psi_D \phi_D)
\label{eq:decay_asym_def}
\eeq
depends implicitly on the abundances of $\psi_D$, $\phi_D$ and $\psi_B$ 
through their chemical potentials - see Eq.~\ref{eq:asym_from_chem_pot}. 
Since the observed baryon asymmetry $n_{\psi_B}/s \sim 10^{-10}$, 
we can restrict our attention to small chemical potentials 
$\mu/T \ll 1$, such that $\Delta \Gamma$ is linear in $\Delta n_i$ to a good approximation.
An analytic expression for $\Delta \Gamma$ can be obtained in the 
interesting limit $\mu/T \ll m_{\phi_D}/T \ll T/M_\varphi \ll 1$\footnote{In
the other limit where final state masses are larger than the temperature, the
resulting asymmetries are Boltzmann suppressed.} 
\beq
\Delta\Gamma 
= \Gamtot \left[
   12 \ln \left[\frac{m^2_{\phi_D}}{2 T^2} \right] \frac{\mu_{\phi_D} T}{M^2_\varphi} 
  - 4\pi^2 \frac{\left(\mu_{\psi_D} +  \mu_{\psi_B} - 4 \mu_{\phi_D} \right)  T^2 }{M_\varphi^3}
\right],
\label{eq:decay_asym_anal}
\eeq
where 
we have assumed that the visible and dark sectors are in thermal 
equilibrium. 
We have kept $\mu_{\phi_D} \neq \mu_{\psi_D}$ to differentiate between the contributions coming from 
final state bosons and fermions.
The logarithm of $m_{\phi_D}/T$ is a remnant of Bose enhancement encoded 
by the stimulated emission factor $(1 + f_{\phi_D})$ in Eq.~\ref{eq:basic_decay_rate}. 
It arises because the phase space density diverges as $E_{\phi_D}\rightarrow \mu_{\phi_D}$ indicating that $\phi_D$ is condensing; 
the divergence is regulated by the mass $m_{\phi_D}$. 
In the left panel of Fig.~\ref{fig:decay_asym_three_body} we 
compare the analytic expression in Eq.~\ref{eq:decay_asym_anal} to 
the numerical evaluation in the limit of small chemical potentials, finding excellent 
agreement in the relevant range of parameters.

As expected, we see that the leading order corrections from Bose enhancement
and Pauli exclusion are of the same sign. This sign is readily understood:
for $\mu_{\psi_D},\; \mu_{\phi_D}>0$ (corresponding to more DM than anti-DM),
Pauli exclusion blocks the channel with $\psi_D$ in the final state, while Bose
enhancement favors the channel involving $\phi_D$. Comparing this with the definition 
of the asymmetry, Eq.~\ref{eq:decay_asym_def}, means that $\Delta\Gamma <0$ for 
$\mu_i > 0$, in agreement with Eq.~\ref{eq:decay_asym_anal}.
Note that the sub-leading correction for Bose enhancement is in
fact larger than the leading order effect from Pauli exclusion for 
$\mu_{\psi_D} = \mu_{\phi_D}$. Therefore, for the rest of the section we will focus on the dominant Bose
enhancement effect. 

The final important feature of Eq.~\ref{eq:decay_asym_anal} is that 
the decay asymmetry is largest at early times and higher $T$. As we show below, 
this causes the bulk of the visible sector asymmetry to be generated 
\emph{well before} the majority of $\varphi$ decays at $t \sim \Gamtot^{-1}$.

\begin{figure}
  {\centering
  \includegraphics[width=0.45\textwidth]{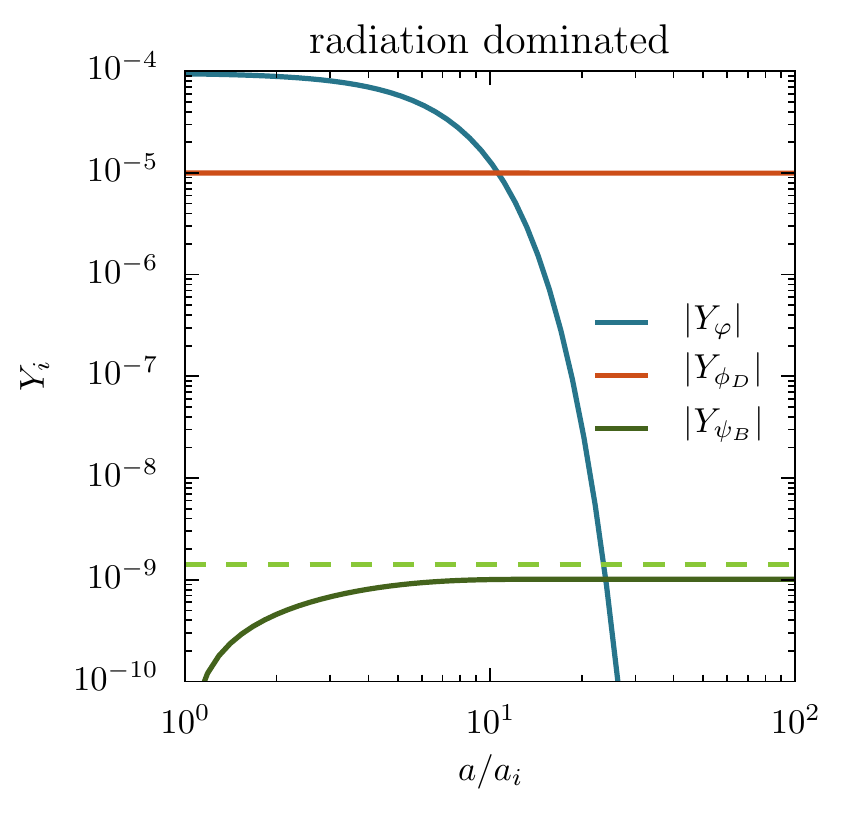}
  \includegraphics[width=0.45\textwidth]{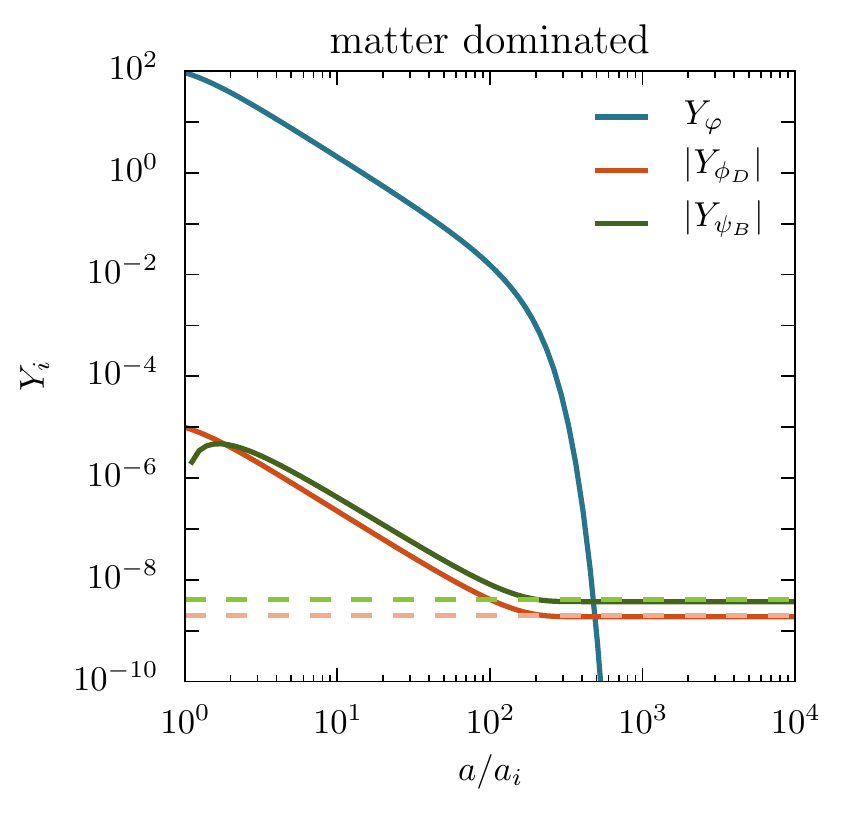}
}
  \caption{Evolution of the baryon and dark matter asymmetries during $\varphi$ decay 
    for radiation-dominated (left plot) and matter-dominated (right plot) 
    initial conditions. We chose $\rho_{\varphi,i}/\rho_{R,i} = 10^{-3}$ in the former case,  
    and $\rho_{\varphi,i}/\rho_{R,i} = 10^{3}$ in the latter. 
    The solid lines show the numerical solution of the Boltzmann equations, while 
    the dashed lines show the analytic approximations for the final abundances discussed in the text. 
    Note that the bulk of the visible asymmetry $Y_{\psi_B}$ is produced before $\varphi$ decays.
    We used $M_\varphi = 100\;\TeV$, $\TRH = 2.5\;\TeV$, $Y_{\phi_D,i} = 10^{-5}$ and 
    $T_i = 3 \TRH$ in these examples. \label{fig:bose_evolution}}
\end{figure}

The system of Boltzmann equations for $\varphi$, $n_{\psi_B}$ (Eqs.~\ref{eq:phi_boltzmann} and 
~\ref{eq:baryon_boltzmann}, respectively) and the DM is closed 
once we include the Friedmann equation
\beq
H^2  = \frac{8\pi}{3\Mpl^2}\left(\rho_R + \rho_\varphi\right),
\eeq
and radiation (or entropy) production due to $\varphi$ decays
\beq
\dot \rho_R + 4H \rho_R = + \Gamtot\rho_\varphi, 
\eeq
where $\rho_\varphi = M_\varphi n_\varphi$. 
The size of the radiation density at the time of the $\varphi$ decay 
determines two distinct possibilities. When $\rho_\varphi \gg \rho_R$, the universe 
is initially $\varphi$-dominated and a large asymmetry produced by the decays is 
diluted by the significant entropy dump. In the opposite regime $\rho_\varphi \ll \rho_R$, 
the universe is radiation dominated. The above system is 
easily solved numerically for any choice of parameters. Sample solutions 
are shown in Fig.~\ref{fig:bose_evolution} for the $\varphi$- and radiation-dominated cases. 
Below we use approximate analytic solutions to better understand these results.

Using the same approximations as before, $\mu/T \ll m_{\phi_D}/T \ll T/M_\varphi \ll 1$, 
we can easily estimate the baryon asymmetry yield.  This limit allows us to neglect
washout reactions generated by the first term in
Eq.~\ref{eq:bose_model_full}, e.g. $\psi_B \psi_D \phi_D^\dagger
\leftrightarrow \psi_B^\dagger \psi_D^\dagger \phi_D$, since these are
suppressed by $(T/M_\varphi)^4 \ll 1$. For the analytic results 
below we make the additional assumption that $\mu_{\phi_D} = \mu_{\psi_D}$, 
i.e. that the transfer reactions $\phi_D \phi_D\leftrightarrow \psi_D\psi_D$ 
are in equilibrium. This ensures that $\Delta n_{\phi_D}a^3$ and $\Delta n_{\psi_D} a^3$ 
are constant. Note that these comoving number densities are insensitive to dilution 
from entropy release.

We first consider the radiation-dominated (RD) scenario where $\rho_\varphi \ll \rho_R \sim T^4$ prior to the decay. 
In this limit, the entropy produced by $\varphi$ is negligible, which means that $Y_{\phi_D} \equiv \Delta n_{\phi_D}/s$ 
is constant. As alluded to above, one can see that the standard intuition of the asymmetry being generated by the decay occurring 
when $\Gamma \sim H$ is incorrect. Comparing $\Delta \Gamma$ to the Hubble rate during radiation domination 
$H \sim T^2/\Mpl \sim 1/t$:
\bea
\Delta\Gamma \sim \Gamtot Y_{\phi_D} \frac{T^2}{M_\varphi^2} \ln(m_{\phi_D}/T) \sim \frac{\ln t}{t}, 
\eea
we find that $\Delta \Gamma/H$ is only logarithmically dependent on time and in fact
favors earlier times!  This means that roughly an equal amount of asymmetry
is being generated every single $e$-folding, suggesting that the naive
instantaneous decay estimate 
must be corrected by the number of $e$-foldings. This is logarithmically sensitive to the 
initial time, which depends on when the out-of-equilibrium $\varphi$ density and 
the dark matter asymmetry were generated.

Using these limits, we can solve the Boltzmann equations analytically.  A
simple closed form can be obtained when $\Gamtot/H_i\ll 1$, where $H_i$ is the initial
Hubble rate that determines the initial time $t_i \sim H_i^{-1}$.
We find the final baryon number abundance to be
\bea
\text{RD:}\qquad \frac{Y_{\psi_B}}{Y_{\phi_D}} &=& k g_* \ln \left[\frac{m^2_{\phi_D}}{2 T^2_i}\right] Y_\varphi 
\left(\frac{\TRH}{M_\varphi}\right)^2 \exp\left(\frac{\Gamtot}{2H_i}\right)
\left|\Ei\left(-\frac{\Gamtot}{2H_i}\right)\right| \nonumber \\
&\approx& k g_* \ln \left[\frac{m^2_{\phi_D}}{2 T^2_i}\right] Y_\varphi 
\left(\frac{\TRH}{M_\varphi}\right)^2 \log\left(\frac{\Gamtot}{2H_i}\right),
\label{eq:rad_dom_asym_anal}
\eea
where $k = 4\pi^2/5 \approx 7.9$, $T_i$ is the initial temperature, $\Ei(-z)$ is the exponential integral with 
$\Ei(-z) \sim \gamma_E + \ln z$ for $z \ll 1$ and in the second line we used the approximation that $\Gamtot \ll H_i$.  
The temperature $\TRH$ is defined
in Eq.~\ref{eq:TRH} and in the radiation-dominated regime it is used as a proxy for $\Gamtot$ rather than 
an actual reheat temperature.  
The correct baryon number asymmetry can be obtained for reasonable values such
as $Y_{\phi_D} \sim Y_\varphi \sim 10^{-4}$ and $M_\varphi \sim 10^{2} \, T$.
This analytical result is compared with the full numerical solution in the left
panel of Fig.~\ref{fig:bose_evolution}.

Next we consider the matter-dominated (MD) case, that is $\rho_\varphi \gg \rho_R \sim T^4$ just before the decay. 
In this case, $\varphi$ decays generate a large amount of entropy, reheating the universe. The reheat temperature 
defined by $\Gamtot = H(\TRH)$ is~\cite{Moroi:1999zb} 
\beq
\label{eq:TRH}
\TRH = \left(\frac{90}{\pi^2 g_*(\TRH)} \right)^{1/4} \sqrt{\Gamma_\varphi \Mpl}.
\eeq

As before, we can see that the asymmetry is produced at early times by comparing $\Delta \Gamma$ to
Hubble $H \sim 1/t$:
\bea
\Delta\Gamma \sim \Gamtot Y_{\phi_D} \frac{T^2}{M_\varphi^2} \ln(m_{\phi_D}/T) \sim \frac{\ln t}{\sqrt{t}}, 
\eea
where we have used $T \sim 1/t^{1/4}$~\cite{Giudice:2000ex}.  From this, we see that $\Delta \Gamma/H$ is 
largest for $t <\Gamtot^{-1}$, so most of the asymmetry
is in fact generated before $\varphi$ decay. 
The later decays are a subdominant contribution to the asymmetry.

\begin{figure}
  {\centering
  \includegraphics[width=0.45\textwidth]{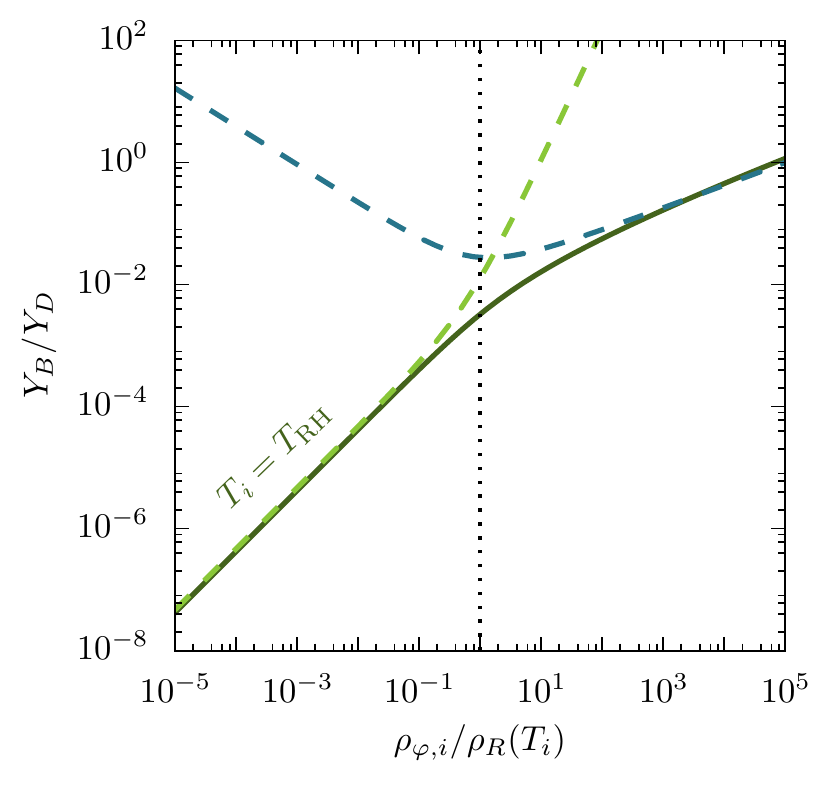}
}
  \caption{Ratio of baryon to DM asymmetry yield at late times as a function of the initial $\varphi$ density. 
    The left (right) dashed lines show the 
    approximate analytic solutions in the radiation-dominated (matter-dominated) universe as discussed in the text.
    We see that these analytic solutions very quickly become good approximations in their respective regions of validity.
    \label{fig:bose_asymmetry_yield_vs_density}}
\end{figure}

The Boltzmann equations 
can be solved for $t \ll \Gamtot^{-1}$ for $\Delta n_{\psi_B}$~\cite{Chung:1998rq,Giudice:2000ex}. The final 
baryon yield $Y_{\psi_B} = \Delta n_{\psi_B}/s_f$ can then be evaluated as
\beq
\text{MD:}\qquad \frac{Y_{\psi_B}}{Y_{\phi_D}} \approx k g_*  \ln \left[\frac{m^2_{\phi_D}}{2 T^2_i}\right]
\left(\frac{\TRH}{M_\varphi}\right)^3 \left(\frac{H_i}{\Gamtot}\right)^{3/4}
,
\label{eq:phi_dom_asym_anal}
\eeq
where we assumed that the logarithmic part of 
$\Delta \Gamma$ is constant. The left hand side contains quantities evaluated at late times; in particular $Y_{\phi_D}$ includes dilution 
due to $\varphi$ decays. The constant $k$ is given by
\beq
k \approx 4 \frac{45}{2}\left(\frac{\pi^2}{30} \right)\left(\frac{2}{5}\right)^{3/4} 
\frac{2\Gamma\left(\frac{9}{20}\right)\Gamma\left(\frac{3}{4}\right)}{\Gamma\left(\frac{1}{5}\right)}
\approx 15.6.
\eeq
Equation~\ref{eq:phi_dom_asym_anal} contains the initial Hubble, indicating that it is a UV dominated process.  
Parametrically $Y_{\psi_B} \lesssim Y_{\phi_D}$ because the last two factors on the right hand side of Eq.~\ref{eq:phi_dom_asym_anal} 
can be written as $(T_{\mathrm{max}}/M_\varphi)^3$, where $T_{\mathrm{max}}$ is the maximum 
temperature achieved during reheating~\cite{Giudice:2000ex}. We require that $T_{\mathrm{max}} < M_\varphi$ 
to avoid washout and to ensure that our approximations for $\Delta \Gamma$ are valid. However,
the baryon asymmetry can be \emph{larger} 
than the DM asymmetry even if $T_{\mathrm{max}} \lesssim M_\varphi$ since $kg_* \sim \mathcal{O}(10^3)$ 
is large for $\TRH$ high enough. This is demonstrated by the numerical solution 
of the Boltzmann equations shown in the right panel of Fig.~\ref{fig:bose_evolution}. For the 
benchmark point shown, $T_{\mathrm{max}}/M_\varphi \lesssim 10^{-1}$ and wash-out is 
expected to be unimportant. In Fig.~\ref{fig:bose_asymmetry_yield_vs_density} we show the baryon yield 
relative to the DM asymmetry for a range of initial $\varphi$ densities interpolating between 
the RD and MD cases discussed above. The analytic solutions presented in Eqs.~\ref{eq:rad_dom_asym_anal} and \ref{eq:phi_dom_asym_anal} 
agree well with the full numerical solution in their regions of validity away from $\rho_{\varphi,i}/\rho_{R}(T_i)\approx 1$.

The above results should be compared with the standard out-of-equilibrium decay scenario where 
$\Delta \Gamma$ does not depend on temperature and the yield $Y_{\psi_B}\approx \Delta\Gamma \, \TRH/(\Gamtot M_\varphi)$ 
is independent of initial conditions~\cite{Kolb:1990vq}. In the case considered in this section, a larger initial $\varphi$ density (larger $H_i$) 
results in the production of a larger asymmetry. Thus we find that when Bose enhancement is responsible for communicating the asymmetry 
between the dark and visible sectors, initial conditions become important. In the following 
section we reach a similar conclusion for the class of models where Pauli blocking rather than Bose 
enhancement is responsible for asymmetry production.
\section{Asymmetries through Pauli Blocking} \label{Sec: pauli}

In the previous section we presented a model where a particle asymmetry was
generated dominantly by Bose enhancement.  In this section we consider the 
complementary case where the leading effect is due to Pauli exclusion. 
This is easily implemented in the toy theory 
\bea
\mathcal{L} = \frac{1}{2} \varphi \bar \psi_B (a + i b \gamma_5 ) \psi_B^C + \lambda \overline{\psi_B^C} \psi_D \Phi_{DB} + \hc
\label{eq:pauli_int}
\eea
where $\psi_B$ (and its charge conjugate $\psi^C_B$) and $\psi_D$ are now Dirac fermions with 
masses $\mB$, $\mD$,
charged under baryon and dark matter number, respectively; $\Phi_{DB}$ is a 
complex scalar with mass $\mDB$, 
carrying opposite charge under both symmetries. Thus, the second term preserves 
both $U(1)_B$ and $U(1)_D$. We will show that decays of the $\varphi$ 
will violate \CP in the presence of a $\psi_D$ asymmetry, even if they are \CP-symmetric 
at zero temperature. As in the previous section, we assume that a $\psi_B$ 
asymmetry can be converted into visible baryons, via, e.g. the 
neutron portal, Eq.~\ref{eq:neutron_portal}.

The decays of $\varphi$ tend to wash out any existing $\psi_B$ asymmetry.
For example, if there are more $\psi_B$
than $\bar \psi_B$ then Pauli blocking biases $\varphi$ decays 
 to generate more $\bar \psi_B$, eventually destroying any baryon number present.
We first consider what happens when there is an non-zero dark matter asymmetry
and zero initial baryon number asymmetry. Conservation of $U(1)_B$ charge ensures that 
\beq
(n_{\psi_B} - n_{\bar \psi_B}) - (n_{\Phi_{DB}} - n_{\Phi_{DB}^\dagger}) = 0,
\label{eq:charge_neutrality}
\eeq
which implies  $\mu_{\psi_B} = \mu_{\Phi_{DB}}$ at temperatures above $\psi_B$ and $\Phi_{DB}$ 
masses. The second interaction in Eq.~\ref{eq:pauli_int} 
enforces chemical equilibrium 
\beq
\mu_{\psi_B} + \mu_{\psi_D} + \mu_{\Phi_{DB}} = 0.
\label{eq:chem_equilibrium}
\eeq
It is easy to solve for
chemical potentials in the limit of small asymmetries, see e.g.~\cite{Harvey:1990qw}; 
the result is
\bea
Y_{\Phi_{DB}} = Y_{\psi_B}  = -\frac{Y_{\psi_D}}{2}, 
\eea
where $Y_i = (n_i - n_{\bar i})/s$. 
Despite the absence of any initial baryon number, $\psi_B$ has a non-zero asymmetry.

Next we consider what happens when $\varphi$ decays. As discussed above 
Pauli exclusion pushes the system towards a
configuration where $Y_{\psi_B} = 0$.  As long as there is a large enough
number density of $\varphi$, $Y_{\psi_B}$ will be driven to zero. 
We can now calculate the baryon number generated by the decays to find
that 
\bea
\label{Eq: baryon number}
Y_B \equiv Y_{\psi_B} - Y_{\Phi_{DB}} =  \frac{Y_{\psi_D} - Y_{\Phi_{DB}}}{2} =  \frac{Y_D}{2} 
\eea
Chemical equilibrium ``hides'' baryon number in the scalar states 
$\Phi_{DB}$, protecting it from wash-out via $\varphi$ decays.
For $\mDB > \mB + \mD$, after $\Phi_{DB}$ freezes
out and decays, its asymmetry will be transferred back into $\psi_B$ and $\psi_D$.
Thus we see that the Pauli exclusion principle can make an otherwise \CP
preserving decay generate baryon number. 

Note that this setup is closely related to models where baryon number
is generated in thermal equilibrium~\cite{Blum:2012nf,Servant:2013uwa}. 
In particular, in Ref.~\cite{Servant:2013uwa} an existing DM asymmetry is used to bias 
electroweak sphalerons (which are in equilibrium prior to the 
electroweak phase transition) to generate a baryon asymmetry. 
A similar scenario is realized in the present model if 
$M_\varphi \lesssim T$ and the relevant couplings 
are large enough, such that $B$-violating scattering 
like $\psi_B\psi_B\leftrightarrow \bar\psi_B\bar\psi_B$ is 
in equilibrium. In this limit one can solve for
the chemical potentials to find the same result of a non-zero baryon
number existing in thermal equilibrium with the value shown in Eq.~\ref{Eq: baryon
number}. 
If $B$-violating $\varphi$-mediated scattering continues after 
$\Phi_{DB}$ freeze-out and decay, 
any existing baryon number would be washed out, so such processes 
must go out of equilibrium. In Ref.~\cite{Servant:2013uwa} 
baryon number violating sphalerons are turned off by a first order electroweak 
phase transition.  In our model, such a rapid shut off is not possible. 
Thus, we focus on 
the out-of-equilibrium decay scenario, where baryon number violation turns off 
once $\varphi$ decays.

An important restriction on this model arises from the fact that the rate at
which baryon number is generated is Boltzmann suppressed in the limit
$M_\varphi \gg T$.  This is
because in the two-body decay $\varphi\rightarrow \psi_B\psi_B$,
the $\psi_B$ final state energy is fixed to be $M_\varphi/2$, while Pauli
exclusion is most effective at energies below the temperature. However,
in the limit where $M_\varphi \ll T$, the inverse decays
$\psi_B\psi_B\rightarrow\varphi$ become important and wash out the asymmetry.
Thus there
is only a finite range of parameters with $M_\varphi \gtrsim \TRH$ where Pauli blocked
decays generate a significant asymmetry. Due to the lack of parametric control, we 
explore this situation numerically and provide a useful 
analytic estimate of the final asymmetry. In the following two subsections, 
we first write down the coupled set of Boltzmann equations and then 
discuss their solutions.

\subsection{Boltzmann Equations}
The Boltzmann equations for the particle asymmetries $\Delta n_i = n_i - n_{\bar i}$ have the form 
\beq
\frac{d \Delta n_i}{dt} + 3H\Delta n_i = C_i[\Delta n_j],
\eeq
where $C_i$ are the collision terms which include the 
effects of number-changing interactions that enforce chemical 
equilibrium. In writing this system of equations we 
approximated the phase space distributions by their Maxwell-Boltzmann 
limits.
For simplicity we make the additional assumption of
\emph{kinetic} equilibrium and small asymmetries, i.e. $\mu_i/T\ll 1$,
such that $n_i + n_{\bar i} \approx 2\neqd_i$. Note that this requires the 
existence of efficient interactions of the $\psi_D$, $\psi_B$ and $\Phi_{DB}$ states 
with the thermal bath, which we leave unspecified. 

Given the interactions in Eq.~\ref{eq:pauli_int},
the $\psi_D$ and $\Phi_{DB}$ collision terms at leading 
order in the coupling $\lambda$ include only $1\leftrightarrow 2 $ processes:
\begin{align}
  C_{\psi_D} & = -\langle \Gamma_{D} \rangle \left(\Delta n_{\psi_D} 
  + \frac{\neqd_{\psi_D}}{\neqd_{\psi_B}} \Delta n_{\psi_B} 
  + \frac{\neqd_{\psi_D}}{\neqd_{\Phi_{DB}}} \Delta n_{\Phi_{DB}}
  \right) + \text{perms.},\\
  C_{\Phi_{DB}} & = -\langle \Gamma_{{DB}} \rangle \left(\Delta n_{\Phi_{DB}} 
  + \frac{\neqd_{\Phi_{DB}}}{\neqd_{\psi_B}}\Delta n_{\psi_B}
  + \frac{\neqd_{\Phi_{DB}}}{\neqd_{\psi_D}}\Delta n_{\psi_D}\right)
  + \text{perms.},
\end{align}
where the $\langle \Gamma_{D} \rangle = \Gamma_{D} K_1(m_{\psi_D}/T)/K_2(m_{\psi_D}/T)$ is the thermally-averaged decay rate 
for $\psi_D\rightarrow \Phi_{DB} \psi_B^\dagger$ and similarly for $\Gamma_{{DB}}$; 
``perms.'' stands for terms with identical structure but with $D$, $B$ and $DB$
permuted. These rates are given in Appendix~\ref{sec:pauli_rates}. For a given
choice of masses, only one of $\Gamma_{D,B,DB}$ is non-zero. 
Note that 
with the above assumptions $\Delta n_i/(2\neqd_i)\approx \mu_i/T$, such that the collision 
terms above vanish when $\mu_{\psi_D} + \mu_{\psi_B} + \mu_{\Phi_{DB}} = 0$, i.e. in chemical equilibrium.

The $\psi_B$ collision term includes additional contributions 
from $B$-violating $\varphi$ decays and $\varphi$ mediated scattering. 
The decay contribution is given by 
\bea
C_{\psi_B} &\supset & \int d\Phi_3 |\mathcal{M}(\varphi\rightarrow \psi_B\psi_B)|^2
\left[f_\varphi (1 - f_{\psi_{B},1})(1-f_{\psi_{B},2}) - f_{\psi_{B},1}f_{\psi_B,2} (1+ f_\varphi) - (\psi_B \rightarrow \bar \psi_B)\right] \nonumber\\
&\approx & - \frac{1}{2M_\varphi}\int d\Phi_2 |\mathcal{M}(\varphi\rightarrow \psi_B\psi_B)|^2  
\left[ 2 n_\varphi \frac{\Delta n_{\psi_B}}{\neqd_{\psi_B}}e^{-M_\varphi/2T}
+ \neqd_\varphi\frac{\left(n_{\psi_B}^2 - n_{\bar \psi_B}^2\right)}{(\neqd_{\psi_B})^2} \right] \nonumber \\
 & = & -2\Gamma_\varphi \left(\frac{\Delta n_{\psi_B}}{\neqd_{\psi_B}}\right)
 \left[n_\varphi e^{-M_\varphi/2T} + \neqd_\varphi\right],
\eea
where in the first step we approximated $E_\varphi\approx M_\varphi$, used detailed balance to replace 
$f_{\psi_B,1} f_{\psi_B,2}$ by $\exp(-E_\varphi/T) n_{\psi_B}^2/(\neqd_{\psi_B})^2$ and performed the integral over $p_\varphi$. The 
former approximation (along with the Maxwell-Boltzmann limit used throughout this section) breaks down 
when $T\sim M_\varphi$. This is also the regime where the DM-induced decay asymmetry 
is largest. Note that in the last line $\Gamma_\varphi$ is the total decay rate (both to $\psi_B\psi_B$ and 
$\bar\psi_B\bar\psi_B$), including a symmetry factor for identical final state particles.
The full collision term can now be written as
\begin{align}
  C_{\psi_B}  = & -\langle \Gamma_{B} \rangle \left(\Delta n_{\psi_B}  
  + \frac{\neqd_{\psi_B}}{\neqd_{\psi_D}} \Delta n_{\psi_D}  
  + \frac{\neqd_{\psi_B}}{\neqd_{\Phi_{DB}}} \Delta n_{\Phi_{DB}}\right) 
  + \text{perms.} \nonumber \\
  & - 2 \Gamma_\varphi \left(\frac{\Delta n_{\psi_B}}{\neqd_{\psi_B}}\right)
 \left(n_\varphi e^{-M_\varphi/2T} + \neqd_\varphi\right) 
 - 4\langle\sigma v\rangle_{\text{RISS}}\neqd_{\psi_B} \Delta n_{\psi_B}, 
 \label{eq:baryon_collision_term}
\end{align}
where $\langle\sigma v\rangle_{\text{RISS}}$ is the Real Intermediate State-subtracted 
(RISS) wash out cross-section for the $\varphi$-mediated process 
$\psi_B\psi_B\leftrightarrow \bar\psi_B \bar\psi_B$ discussed in 
Appendix~\ref{sec:RIS}.
Using the same approximations we can write the $\varphi$ collision term as:
\beq
C_\varphi = - \Gamma_\varphi\left(n_\varphi \left[1 - 2 e^{-M_\varphi/(2T)}\right] - 
\neqd_\varphi \right).
\label{eq:phi_collision_term}
\eeq
The remaining Boltzmann equation governs the radiation density 
\beq
\dot \rho_R + 4H \rho_R = - M_\varphi C_\varphi, 
\eeq
where the minus on the right-hand side simply cancels the one in Eq.~\ref{eq:phi_collision_term}.
In the following section we study this system of Boltzmann equations analytically and numerically.

\subsection{Solutions and Numerical Examples\label{sec:pauli_num_ex}}
Simple approximations to the Boltzmann equations discussed in the previous 
section allow us to determine the final baryon asymmetry analytically. 
For concreteness we assume that the universe is radiation dominated with an 
out-of-equilibrium $\varphi$ density.\footnote{In a $\varphi$ dominated universe 
$\varphi$ decays deposit a large amount of entropy, diluting 
the newly-created baryon asymmetry. Such an entropy release 
can be compensated for with a higher initial DM asymmetry.}
The $\varphi$ equation can be integrated neglecting 
inverse processes and statistical factors in the collision term of Eq.~\ref{eq:phi_collision_term}.\footnote{The 
  omission of the small chemical potentials in this step corresponds 
  to dropping sub-leading $\mathcal{O}(\mu/T)$ terms in the final asymmetry.}
Inserting the resulting solution into the Eq.~\ref{eq:baryon_collision_term}, 
assuming chemical potentials are small and integrating 
this equation we find the baryon asymmetry 
\beq
Y_{\psi_B} = \frac{\Delta n_{\psi_B}}{s} \approx - 4 \left(\frac{\mu_{\psi_B,i}}{T_i}\right) \left(\frac{\Gamma_\varphi}{H_i}\right)
\left[Y_{\varphi,i}f_1(M_\varphi/T_i,\Gamma_\varphi/H_i) + Y^{\mathrm{eq}}_{\varphi,i} f_2(M_\varphi/T_i)\right]
\label{eq:pauli_anal_sol}
\eeq
where quantities with the subscript $i$ are evaluated at the initial time and temperature. In particular 
$\mu_{\psi_B,i}/{T_i}$ can be determined from the DM asymmetry using the charge neutrality and chemical equilibrium 
conditions, Eqs.~\ref{eq:charge_neutrality} and~\ref{eq:chem_equilibrium}, respectively:
\beq
\frac{\mu_{\psi_B,i}}{T_i} = - \frac{\mu_{\psi_D,i}}{T_i} \begin{cases}
  1/2& T_i \gg m \\
  (1 + \neqd_{\psi_B}/\neqd_{\Phi_{DB}})^{-1} & T_i < m.
\end{cases}
\eeq
Finally, the functions $f_1$ and $f_2$ have the following form in the limit $ z = M_\varphi/T_i \gg 1$ and $\gamma = \Gamma_\varphi/H_i\ll 1$:
\bea
f_1 (z,\gamma) & \sim & 2 z^{-4}\left(2 z^2 + z^3 - \gamma \left[24 + 12 z + 2 z^2\right]+ \mathcal{O}(z\gamma^2)\right) e^{-z/2}, \\ 
f_2 (z) & \sim & z^{-1} + \frac{5}{2} z^{-2} + \frac{15}{4} z^{-3} + \mathcal{O}(z^{-4}).
\eea
Note that $Y^{\mathrm{eq}}_{\varphi,i} \sim z^{3/2} \exp(-z)$ so that both
terms in Eq.~\ref{eq:pauli_anal_sol} are Boltzmann suppressed in the
non-relativistic limit. As in the Bose case considered in Sec.~\ref{Sec: bose}, 
we see that the asymmetry production favors early times and higher temperatures. 
In fact, the bulk of the asymmetry is produced before $t\sim \Gamma_\varphi$.
Note that this only holds up to $T_i \sim M_\varphi$ where the 
asymmetry would be damped by the (neglected) wash-out terms. 

The sensitivity of the final asymmetry to early times emphasizes the importance of initial conditions 
in this scenario. In particular, a physical 
set of initial conditions depends on the origins of the $\varphi$ density and DM 
asymmetry. There are several ways to obtain an out-of-equilibrium density 
of $\varphi$. The simplest mechanism for this is freeze-out, which would occur at $T\sim M_\varphi/20$.
Thus the decay and asymmetry production would happen at even lower temperatures. 
From the analytical solution, Eq.~\ref{eq:pauli_anal_sol}, it is clear that 
the final asymmetry would be exponentially suppressed.

Another possibility for generating an out-of-equilibrium density of $\varphi$ is 
the misalignment mechanism. 
If $\varphi$ was displaced from the minimum 
of its potential during inflation, then its field value would remain Hubble damped 
until $H\sim M_\varphi$ (corresponding to a temperature 
$T_{\mathrm{osc}} \sim \sqrt{M_\varphi \Mpl}$). 
At this point $\varphi$ begins coherent oscillations, with 
the energy density in these fluctuations red-shifting as matter. The asymmetry 
generation cannot begin until the DM develops a chemical potential, since it is required to 
bias $\varphi$ decays. Thus, in principle, the $\psi_B$ asymmetry 
can be created any time between $T_{\mathrm{osc}}$ and $\TRH\sim \sqrt{\Gamma_\varphi \Mpl}$.
However, as discussed above, we are working in the Maxwell-Boltzmann limit, 
so our Boltzmann equations are valid only for $T < M_\varphi$.
Thus we confine our attention to the region of parameters where $\TRH < M_\varphi \ll T_{\mathrm{osc}}$.
We emphasize that this is not a fundamental requirement, but merely a computational aid.
The condition $\TRH < M_\varphi$ has the additional benefit of making 
the rates for wash out processes, i.e., inverse decays and $B$-violating $2\rightarrow 2$ 
scattering, very slow.

\begin{figure}
  {\centering
  \includegraphics[scale=0.85]{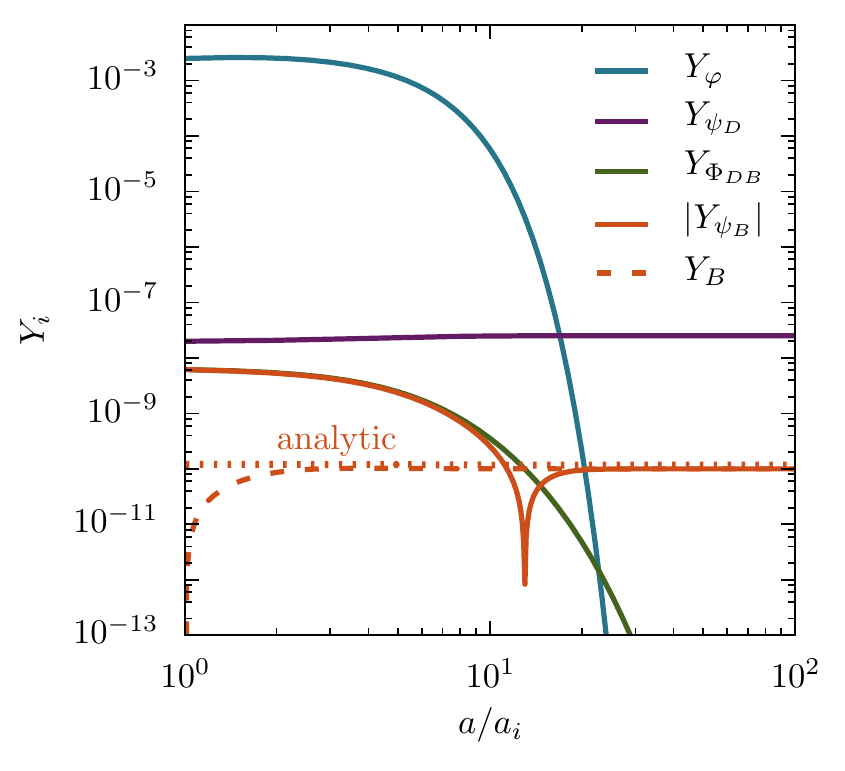}
  \includegraphics[scale=0.85]{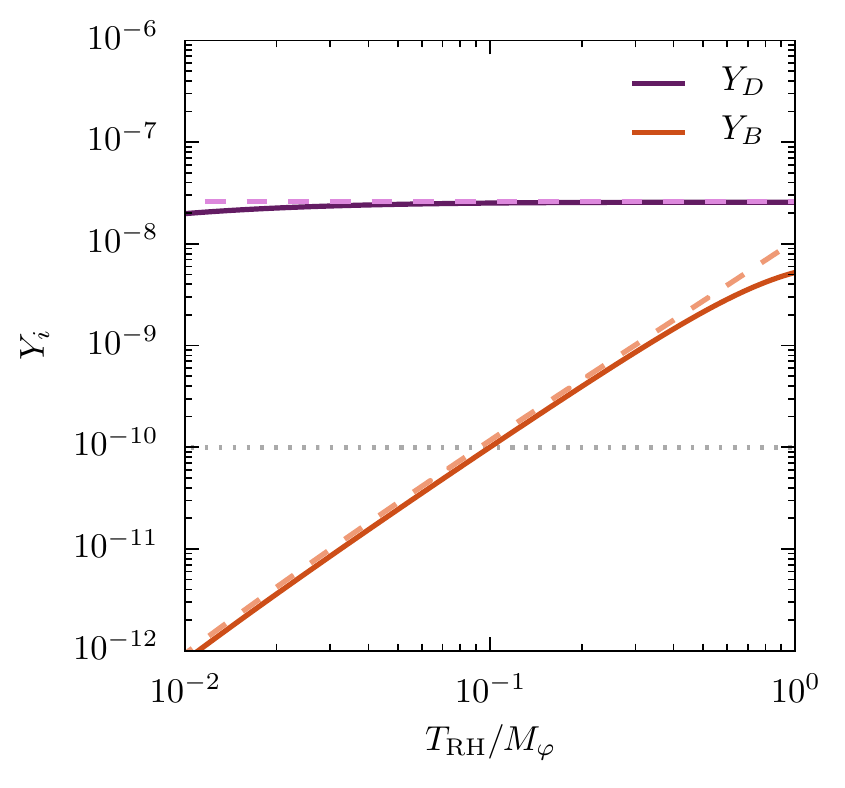}
  }
  \caption{ Numerical solutions to the Boltzmann equations describing 
    asymmetry production through Pauli-blocked decays of $\varphi$. 
    The left panel shows the evolution of the various number densities 
    as a function of the scale factor. The dotted line is the analytic 
    approximation of Eq.~\ref{eq:pauli_anal_sol}. The dashed  
    line shows that the net baryon number, $Y_B=Y_{\psi_B} - Y_{\Phi_{DB}}$ is 
    produced at early times, before $\varphi$ decays.
    The right panel shows the dependence of the final 
    baryon yield on the decay temperature of $\varphi$, $\TRH$.
    The red dashed line is the analytic solution for $Y_{\psi_B}$, while 
    the purple dashed line is the initial DM asymmetry. This initial 
    asymmetry is diluted by the $\varphi$ entropy injection at low $\TRH$.
    The initial conditions and parameter values used are described in the text.
  \label{fig:numsol}}
\end{figure}

We show the numerical
solutions to the system of Boltzmann equations in Fig.~\ref{fig:numsol} for
$M_\varphi = 1\;\TeV$, $\TRH = 100\;\GeV$, 
$T_i = M_\varphi/3$ and $\rho_\varphi(T_i) = 10^{-2} \rho_R(T_i)$.
The initial DM asymmetry is chosen to be $Y_{\psi_D}/s = 2\times 10^{-8}$ in order to 
obtain $Y_B = 10^{-10}$ at late times; 
the other asymmetries are determined from the initial $U(1)_B$ 
charge neutrality and chemical equilibrium requirements in Eqs.~\ref{eq:charge_neutrality} 
and~\ref{eq:chem_equilibrium}, respectively.
The remaining masses are chosen to satisfy 
$M_\varphi > 2\mB$ and $\mDB > \mD + \mB$:
$\mB = 100\;\GeV$, $\mD = 150\;\GeV$ and $\mDB = 300\;\GeV$.
Note, however, that the mechanism is not sensitive to a particular 
choice of masses, as long as the relevant processes are kinematically allowed.
These initial conditions are chosen to produce the correct order 
of magnitude for the baryon asymmetry. 

In the left panel of Fig.~\ref{fig:numsol} we show the evolution of 
various number densities as a function of the scale factor $a$, 
normalized to $a_i$, its value at the initial time.
At early times, chemical equilibrium with the asymmetric 
DM results in non-zero asymmetries for $\psi_B$ and $\Phi_{DB}$, 
while $Y_{\Phi_{DB}} = Y_{\psi_B}$ is guaranteed due to vanishing initial $U(1)_B$ charge. However, $\varphi$ 
decays quickly drive the $Y_{\psi_B}$ asymmetry to $0$, such that the only remaining 
$B$ number is stored in $\Phi_{DB}$. When $\Phi_{DB}$ decays, its asymmetry flows 
into $\psi_B$ and $\psi_D$, resulting in a final non-zero $B$ number. The 
dashed line in this figure shows the net $B$ number density, i.e. 
$Y_{\psi_B} - Y_{\Phi_{DB}}$, which explicitly shows that the $B$ asymmetry
is generated well before $\varphi$ fully decays. The dotted line shows 
the analytic solution of Eq.~\ref{eq:pauli_anal_sol}.

In the right panel of Fig.~\ref{fig:numsol} we show the final 
DM and baryon asymmetries as a function of $\TRH$. The light red dashed 
line shows the analytic solution, while the dashed purple line 
shows the initial DM asymmetry. Increasing $\TRH$ corresponds to 
higher decay rates $\Gamma_\varphi$, which in turn, leads to more $\varphi$ 
decaying at early times, thereby enhancing the effect of Pauli blocking 
and asymmetry generation. Note that this cannot continue 
indefinitely, because at a high enough temperature wash out due to 
inverse decays and $2\leftrightarrow2$ reactions becomes important.
We do not extend the calculation to $\TRH/M_\varphi > 1$ because 
the Maxwell-Boltzmann approximation used throughout this 
section breaks down, but as emphasized before, this is not a 
fundamental limitation of this scenario. Note that at low 
$\TRH$ the entropy injection due to $\varphi$ decays is large enough 
to partially dilute the initial DM asymmetry, resulting in a decreased 
$Y_D$ yield.

\section{Discussion and Conclusion} \label{Sec: conclusion}

In this paper we have investigated \CP violation through matter effects at 
finite temperature in the early universe. We have shown that 
quantum statistical effects play a crucial role 
in this class of baryogenesis models. We considered 
models where the visible sector Lagrangian is \CP invariant and \CP violation 
arises from an asymmetric background density of particles. Such a background is expected to exist, 
if, e.g., dark matter is asymmetric or if there is another quasi-stable asymmetric 
species during this epoch.
Out-of-equilibrium baryon number violating decays of a scalar $\varphi$ were shown to generate 
baryon number in this background, despite the absence of \CP violation 
in the interactions of $\varphi$. This is distinct from the standard out-of-equilibrium 
decay scenario employed in, e.g., leptogenesis or GUT baryogenesis, where 
\CP is violated through the interference of tree and loop contributions to the 
decay rate.
Instead, the dark matter asymmetry biases $\varphi$ decay to prefer some channels over others through Pauli blocking or Bose
enhancement of the corresponding final states. We considered two toy models where 
only one of these effects is dominant. 
Thus, we provide the first example of a baryogenesis scenario where
statistical factors play a critical role -- no asymmetry is produced 
when Pauli blocking and Bose enhancement are neglected.

Finally, we note that the models considered in this paper are far from
complete. While they illustrate the general ``\CP violation through matter
effects'' mechanism, we have not attempted to embed them in realistic scenarios,
which, e.g., describe the $\varphi$ production mechanism in the early universe,
the origin of baryon number violation, or the nature of the dark sector
asymmetry. These details are important for setting the initial conditions 
which play a crucial role in determining the baryon yield 
when the $B$-violating decay asymmetry is generated by quantum statistical effects.
Moreover, these directions may identify concrete experimental probes 
for this class of models. A complete
model of baryon number violation in our scenarios (i.e. a UV completion of the
non-renormalizable operators in Eqs.~\ref{eq:bose_model_full} and
~\ref{eq:neutron_portal}) may be testable at nucleon decay or neutron
oscillation experiments.  Similarly, different mechanisms of generating the
dark matter asymmetry can give rise to observable signatures. For example, if
the dark asymmetry is created in a strongly first order phase transition \`a la
electroweak baryogenesis, a detectable gravitational wave background may be
generated. The necessity for interactions between the dark and visible sectors
in our scenarios suggests the possibility of complementary probes. We leave a
detailed investigation of these issues to future work.

\acknowledgments
We thank Brian Shuve and Falko Dulat for useful conversations and Yanou Cui for stimulating
discussions that led to this work.
N.B. is supported by DOE Contract DE-AC02-76SF00515. 
A.H. is supported by the DOE Grant DE-SC0012012 and NSF Grant 1316699.

\appendix
\section{Wash-out and Decay Rates in the Pauli Scenario\label{sec:pauli_rates}} 
In this Appendix we collect the decay rates and wash-out cross-section 
relevant for the scenario in discussed in Sec.~\ref{Sec: pauli}.

\subsection{Decays}
The first term in Eq.~\ref{eq:pauli_int} gives rise to $B$-violating $\varphi$ decays 
with the rate
\beq
\Gamma (\varphi \rightarrow \psi_B\psi_B) = \Gamma (\varphi \rightarrow \bar \psi_B \bar \psi_B)
= \frac{M_\varphi}{32\pi}
\left[
  |a|^2 + |b|^2 - 4 y^2 |a|^2
\right]
\left[
  1- 4y^2
\right]^{1/2},
\eeq
where $y = \mB/M_\varphi$. Assuming these are the only two channels the total 
$\varphi$ decay rate is 
\beq
\Gamma_\varphi
= \frac{M_\varphi}{16\pi}
\left[
  |a|^2 + |b|^2 - 4 y^2 |a|^2
\right]
\left[
  1- 4y^2
\right]^{1/2}.
\eeq

The second term in Eq.~\ref{eq:pauli_int} is responsible for sharing the $U(1)_B$ and $U(1)_D$ 
numbers, such that $\varphi$ decays generate a baryon asymmetry. 
The leading physical processes that enforce this are the following decays and inverse decays 
\begin{align}
  \Gamma_{DB} = \Gamma(\Phi_{DB}\rightarrow \bar \psi_D \bar \psi_B) = \Gamma(\Phi_{DB}^\dagger \rightarrow  \psi_D \psi_B) & = \frac{|\lambda|^2}{8\pi}\mDB \beta(\mDB, \mD, \mB), \\
  \Gamma_{B} = \Gamma(\psi_B \rightarrow \bar \psi_D \Phi_{DB}^\dagger) = \Gamma(\bar \psi_B\rightarrow \psi_D \Phi_{DB}) & = \frac{|\lambda|^2}{16\pi}\mB \beta(\mB, \mD, \mDB), \\
  \Gamma_D = \Gamma(\psi_D\rightarrow  \Phi_{DB}^\dagger \bar \psi_B)= \Gamma(\bar \psi_D\rightarrow  \Phi_{DB}  \psi_B) & = \frac{|\lambda|^2}{16\pi}\mD \beta(\mD, \mDB, \mB),
\end{align}
where 
\beq
\beta(x,y,z) = \frac{|\mDB^2-(\mB+\mD)^2|}{x^2}\left[1-\frac{(y+z)^2}{x^2}\right]^{1/2}\left[1-\frac{(y-z)^2}{x^2}\right]^{1/2} \theta (x-y-z).
\eeq
Note that for a given mass ordering only one of these rates is non-zero.

\subsection{Scattering and Real Intermediate States\label{sec:RIS}}
In addition to $\varphi$ decays, the $B$-violating interaction in Eq.~\ref{eq:pauli_int} 
induces the wash-out reaction $\psi_B \psi_B\leftrightarrow \bar \psi_{B} \bar \psi_{B}$ with an 
intermediate $s$-channel $\varphi$. If this process is 
active it will drive any existing baryon asymmetry to zero. Fortunately, 
the rate for this wash out process is small when $\varphi$ decays out-of-equilibrium, 
as required in the scenario of Sec.~\ref{Sec: pauli}. However, 
this rate can be enhanced when the intermediate $\varphi$ goes on-shell, so 
we evaluate it below.
The corresponding cross-section is 
\begin{align}
\sigma v_{\mathrm{lab}} & = 
\frac{1}{128\pi s} \frac{\left(1-4\mB^2/s\right)^{1/2}}{\left(1-2\mB^2/s\right)\left(1-M_\varphi^2/s\right)^2}
\left(
|a|^2 + |b|^2 - \frac{4\mB^2}{s} |a|^2
\right)^2 \\
& \approx \frac{|b|^4 \mB^2}{16\pi M_\varphi^4\left(1-4\mB^2/M_\varphi^2\right)^2}\sqrt{\epsilon} + \mathcal{O}(\epsilon^{3/2}),
\end{align}
where $\epsilon = (s-4\mB^2)/4\mB^2$ is the kinetic energy 
per unit mass in the lab frame~\cite{Gondolo:1990dk,Edsjo:1997bg}. 
Note that the leading 
contribution from the scalar interaction $\propto a$ is velocity 
suppressed. This is because the bilinear $\bar \psi_B \psi_B^C$ can only 
create states with orbital angular momentum $L=1$~\cite{Kumar:2013iva}. 
The thermal average can be performed analytically in the 
large $x=\mB/T$ limit or numerically, keeping all $\epsilon$ 
and $x$ dependence in the cross-section. The analytical 
result is 
\beq
\langle\sigma v\rangle (\psi_B \psi_B\leftrightarrow \bar \psi_B\bar \psi_B) 
= \frac{1}{\sqrt{\pi x}}
\left(\frac{|b|^4 \mB^2}{16\pi M_\varphi^4\left(1-4\mB^2/M_\varphi^2\right)^2} \right),
\label{eq:largex_xsec}
\eeq
where the quantity in the braces is the cross-section 
at threshold. We are interested in this rate while $\psi_B$ and $\bar\psi_B$ 
are still in chemical equilibrium, which means $x$ is not large. 
Moreover, for long-lived $\varphi$ the cross-section is 
strongly peaked around the resonance, away from $\epsilon = 0$. 
This means that the $\epsilon$ expansion is not 
valid. The full rate (without making these approximations) 
is then determined by numerically performing the integral\footnote{This thermal average procedure is valid only in the non-relativistic limit, 
so the results for $x\lesssim 1$ should be considered to be rough estimates.}
\beq
\langle\sigma v\rangle (\psi_B \psi_B\leftrightarrow \bar \psi_B\bar \psi_B)
= \frac{x}{K_2(x)^2} 
\int d\epsilon \sqrt{\epsilon} (1+2\epsilon) K_1(2x\sqrt{1+\epsilon})
\sigma v_{\mathrm{lab}},
\label{eq:sigv_full}
\eeq
where we included a factor of $1/2$ for identical initial states in the 
definition of the thermal average.
This rate includes processes occurring through on-shell 
$\varphi$ exchange when $s = M^2_\varphi$.
However, the on-shell decays and inverse decays 
$\varphi \leftrightarrow \psi_B \psi_B,\;\bar \psi_B\bar \psi_B$ are already 
present in the Boltzmann equations -- see Eq.~\ref{eq:baryon_collision_term}. To avoid double counting the 
resonant contribution to 
$\langle\sigma v\rangle (\psi_B \psi_B\leftrightarrow \bar \psi_B\bar \psi_B)$ 
must be subtracted~\cite{Kolb:1979qa,Giudice:2003jh,Pilaftsis:2003gt}. One simple approach 
to implement this Real Intermediate State (RIS) subtraction 
is to modify the squared $s$ channel propagator as 
\beq
\frac{1}{(s-M_\varphi^2)^2 + \Gamma_\varphi ^2 M_\varphi^2} 
\rightarrow
\frac{1}{(s-M_\varphi^2)^2 + \Gamma_\varphi ^2 M_\varphi^2} 
- \frac{\pi}{M_\varphi \Gamma_\varphi} \delta (s - M^2_\varphi),
\eeq
where $\Gamma_\varphi$ is the total decay rate. 
\begin{figure}[!t]
  \centering
  \includegraphics[scale=0.85]{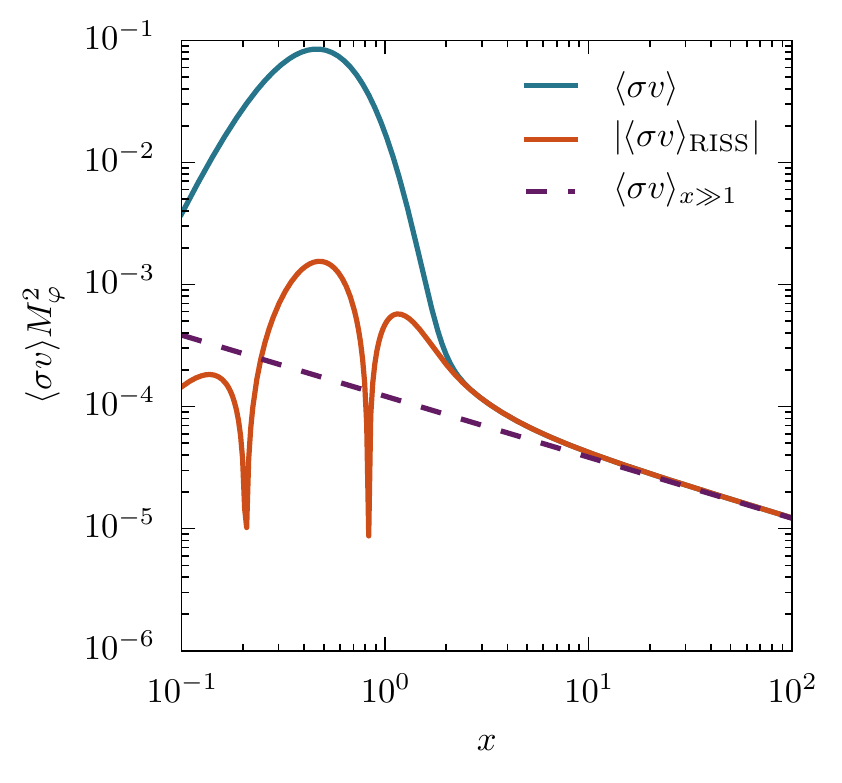}
  \caption{Thermally averaged cross-section for $\psi_B \psi_B\leftrightarrow \bar \psi_B\bar \psi_B$ as a function of 
    $x=\mB/T$ for $a = 0$, $b = 1$ and $\mB/M_\varphi = 0.1$. The solid blue line is the standard thermal average computed from the numerical integral in Eq.~\ref{eq:sigv_full}, which includes the resonant enhancement for $s\approx M_\varphi^2$. The solid orange line shows 
    same cross-section with the real intermediate state contribution subtracted to avoid double counting in solving the Boltzmann equations. 
    The dashed line is the analytical large $x$ approximation from Eq.~\ref{eq:largex_xsec}. 
    \label{fig:xsec_compare}}
\end{figure}
The RIS contribution corresponding to the Dirac delta function 
is easily computed:
\begin{align}
\langle\sigma v\rangle_{\text{RIS}} & = 
\frac{1}{2M^2_\varphi}\frac{\pi^2 x K_1(x/y)}{y^5 K_2(x)^2} 
\mathrm{Br}(\varphi\rightarrow \psi_B \psi_B)
\mathrm{Br}(\varphi\rightarrow \bar \psi_B\bar \psi_B)\frac{\Gamma_\varphi}{M_\varphi} \\
& = \frac{n_\varphi^{\mathrm{eq}}}{(n_{\psi_B}^\mathrm{eq})^2}
\mathrm{Br}(\varphi\rightarrow \psi_B\psi_B)
\mathrm{Br}(\varphi\rightarrow \bar \psi_B\bar \psi_B)\frac{K_1(x/y)}{K_2(x/y)} \Gamma_\varphi.
\label{eq:sigv_ris}
\end{align}
In the last line we wrote the RIS rate in terms of the equilibrium distributions 
for $\psi_B$ and $\varphi$. The proper RIS-subtracted (RISS) rate that 
enters the Boltzmann equation~\ref{eq:baryon_collision_term} is obtained by taking the difference of Eqs.~\ref{eq:sigv_full} and 
~\ref{eq:sigv_ris}.
We compare the RIS-subtracted cross-section with other approximations in Fig.~\ref{fig:xsec_compare}. 
Note that the RISS cross-section becomes negative in the resonance region 
near $x\sim 1$. Numerically this happens because the RIS rate in Eq.~\ref{eq:sigv_ris} is slightly 
larger than the standard rate in Eq.~\ref{eq:sigv_full}; this, in turn, is because of the finite 
width of the resonance peak. This was also observed in Ref.~\cite{Cline:1993bd}.

%

\end{document}